\documentclass[orivec,envcountsame,a4paper]{llncs} 

\usepackage{amsmath}
\usepackage{amssymb}
\usepackage{booktabs}
\usepackage{fancyvrb}
\usepackage{graphics}
\usepackage{microtype}  
\usepackage{multirow}
\usepackage{paralist}
\usepackage{pygmentize}
\usepackage{rgalg}
\usepackage{times}
\usepackage{wrapfig}
\usepackage{xcolor}
\usepackage{xspace}

\usepackage{tikz}
\usetikzlibrary{arrows,positioning,fit,backgrounds,shadows,calc}
\tikzset{
    >=stealth',
    punkt/.style={
           rectangle,
           rounded corners,
           draw=black, thick,
           text width=11.5em,
           minimum height=2em,
           text centered},
    pil/.style={
           ->,
           shorten <=2pt,
           shorten >=2pt,},
    qwe/.style={
           -,
           shorten <=2pt,
           shorten >=2pt,}           
}

\usepackage[colorlinks]{hyperref} 
\RecustomVerbatimEnvironment{Verbatim}{BVerbatim}{}
\definecolor{darkred}{rgb}{0.4,0,0}
\definecolor{darkblue}{rgb}{0,0,0.4}
\definecolor{verylightgray}{rgb}{0.9,0.9,0.9}
\definecolor{lightblue}{rgb}{0,0,0.9}
\hypersetup{colorlinks,linkcolor=darkblue,citecolor=darkblue,urlcolor=darkblue}
\hypersetup{
  pdftitle={Runtime Verification Based on Register Automata},
  pdfauthor={Dino Distefano and Radu Grigore and Rasmus Lechedahl Petersen and Nikos Tzevelekos}}

\title{Runtime Verification Based on Register Automata}
\author
  {     Radu Grigore\inst{1}
  \and  Dino Distefano\inst{1}
  \and  Rasmus Lerchedahl Petersen\inst{2}
  \and  Nikos Tzevelekos\inst{1} }
\institute{
        $\text{}$\inst{1}Queen Mary University of London \
  and \ $\text{}$\inst{2}Microsoft Research}

\newcommand\boldemph[1]{\textbf{\textit{#1}}}

\newcommand{\noterg}[2]{\marginpar{\parbox{31mm}{\raggedright\scriptsize \textcolor{red}{#1}: #2}}}
\newcommand{\rg}[1]{\noterg{rg}{#1}}

\newcommand{\dd}[1]{\noterg{dd}{#1}}

\newcommand{\nt}[1]{{\color{blue}#1}}
\renewcommand{\nt}[1]{#1}

\newcommand{\A}{\ensuremath{\mathcal{A}}}

\newcommand{\error}{\ensuremath{\textcolor{darkred}{\mathtt{error}}}\xspace}

\newcommand{\pattern}[1]{\ensuremath{\mathtt{\underline{#1}}}}
\newcommand{\pmap}{\rightharpoonup}

\newcommand{\start}{\ensuremath{\mathtt{start}}\xspace}
\newcommand{\verbline}[2][]{\[\text{\Verb@#2@}#1\]}

\newcommand{\quoteindent}{1.5\parindent} 
\newcommand{\eqquote}[2]{{%
  \refstepcounter{equation}\label{#2}%
  \newdimen\qi\qi=\quoteindent
  \setbox0=\vbox{\advance\hsize by-2\qi\noindent\em#1}%
  \newdimen\x\x=\ht0 \advance\x by\dp0%
  \setbox1=\vbox to\x{\vss\hbox{(\arabic{equation})}\vss}%
  \leavevmode\\[1ex]%
  \hbox to\hsize{\hskip\qi\box0\hfil\box1}%
  \\[1ex]}}
\newcommand{\eqquotee}[2]{{%
  \refstepcounter{equation}\label{#2}%
  \newdimen\qi\qi=\quoteindent
  \setbox0=\vbox{\advance\hsize by-2\qi\noindent\em#1}%
  \newdimen\x\x=\ht0 \advance\x by\dp0%
  \setbox1=\vbox to\x{\vss\hbox{(\arabic{equation})}\vss}%
  \leavevmode\\[1ex]%
  \hbox to\hsize{\hskip\qi\box0\hfil\box1}%
}}
\newcommand{\emquote}[1]{{%
  \\[1ex]%
  \newdimen\qi\qi=\quoteindent%
  \hbox to\hsize{\hfil\vbox{\advance\hsize by-2\qi\noindent\em#1}\hfil}%
  \\[1ex]}}

\makeatletter
\newenvironment{oneshot}[1]{\@begintheorem{#1}{\unskip}}{\@endtheorem}
\makeatother

\renewcommand{\noterg}[2]{}

\begin{document}
\maketitle
\begin{abstract} 
We propose TOPL automata as a new method for runtime verification of systems
with unbounded resource generation.
Paradigmatic such systems are object-oriented programs which can dynamically 
generate an unbounded number of fresh object identities during their execution.
Our formalism is based on register automata, a particularly successful 
approach in automata over infinite alphabets which administers a finite-state machine with boundedly many input-storing registers.
We show that TOPL automata are equally expressive to register automata 
and yet suitable to express properties of programs.
Compared to other runtime verification methods, our technique can 
handle a class of properties beyond the reach of current tools.
We show in particular that properties which require value updates are not 
expressible with current techniques yet are naturally captured by TOPL 
machines.
On the practical side, we present a tool for runtime verification of 
Java programs via TOPL properties, where the trade-off between the 
coverage and the overhead of the monitoring system is tunable by means 
of a number of parameters.
We validate our technique by checking properties involving multiple 
objects and chaining of values on large open source projects.
\end{abstract}

\section{Introduction}

Runtime verification~\cite{dblp:conf/kbse/havelundr01,dblp:journals/jlp/leuckers09} connotes the monitoring of program executions in order to detect specific error traces which correspond to violations of sought safety properties.
In contrast to its static counterpart, runtime verification checks only certain program executions, yet the error reports are accurate as detected violations represent real bugs in the program.
In the case of systems with dynamic generation of resources, such as object references in Java, runtime verification faces the key challenge of reasoning about a potentially {\em unbounded} number of parameter values representing resource identities.
Hence, the techniques applicable in this realm of programs must be able to deal with infinite alphabets (this idiom is also known as parametric monitoring).
\nt{Leading runtime verification techniques tackle the issue using different approaches, such as reducing the problem to checking projections of execution traces over bounded sets of data values (\emph{trace slicing})~\cite{dblp:journals/corr/abs-1112-5761,dblp:conf/icse/jinmlr12,arnold:2008,dblp:conf/oopsla/allanachklmsst05}, or employing abstract machines whose transition rules are explicitly parameterised~\cite{dblp:journals/logcom/barringerrh10,barringer2010formal,dblp:conf/fm/barringerh11}.}

Another community particularly interested in reasoning over similar data domains, albeit motivated by XML reasoning and model-checking, is the one working on
automata over infinite alphabets. 
Their research has been prolific in developing a wide range of paradigms and accompanying logics, with varying degrees of expressivity and effectiveness (see~\cite{dblp:conf/csl/segoufin06} for an overview from 2006).
A highly successful such paradigm is that of
\emph{Register Automata}~\cite{dblp:journals/tcs/kaminskif94,dblp:conf/mfcs/nevensv01}, which are finite-state machines equipped with a fixed number of registers where input values can be stored, updated and compared with subsequent inputs.
They provide a powerful device for reasoning about temporal relations between a possibly unbounded number of objects in a finite manner. 
In this work we propose a foundational runtime verification method based on a novel class of machines called \emph{TOPL automata}, which connects the field with the literature on automata over infinite alphabets and, more specifically, with register automata.

The key features of our machines are: \begin{inparaenum}[(1)] 
\item the use of registers, and
\item the use of sets of active states (non-determinism).
\end{inparaenum} 
From the verification point of view, registers allow us to use a fixed amount of specification variables which, however, can be \emph{re-bound} (i.e.~have their values updated).
On the other hand, by being able to spawn several active states, we can select different parts of the same run to be stored and processed. 
These features give us the expressive power to capture a wide range of realistic program properties in a {\em finite} way.
A specific such class of properties concerns \emph{chaining} or \emph{propagation}, which are of focal importance in areas like dynamic taint analysis~\cite{dblp:conf/ndss/newsomes05} as well as dynamic shape analysis.\footnote{Although shape analysis is mainly a static technique, we will see in Section~\ref{sec:motivating-examples} that, when doing run-time monitoring,
being able to reason about shapes may be vital.}
\rg{I would remove the footnote.[dd: people may get confusing here if we say shape analysis as it's a static analysis. That's why I want to specify] }
In the latter case, we aspire to reason at runtime about particular shapes of dynamically allocated data-structures irrespectively of their size. 
For example, checking
\eqquote{``the shape of the list should not contain cycles"}{q:concur-it4}
for lists of any size and in a finite way, requires two activities. First, being able to change the value of the variables in the specification while traversing the list (re-binding). Second, keeping  correlations of different elements in the list at the same time (multiple active states). 

\begin{wrapfigure}{r}{.35\textwidth}\centering\small
\vspace{-7mm}
\begin{tikzpicture}[auto,node distance=5mm]
  \node[punkt] (lang) {TOPL Properties};
  \node[punkt] (rtopl) [below=of lang] {hl-TOPL Automata};
  \node[punkt] (topl) [below=of rtopl] {TOPL Automata};
  \node[punkt] (ra) [below=of topl] {Register Automata};
  \path[draw,->,>=latex]
    (lang) edge (rtopl)
    (rtopl) edge[<->] (topl)
    (topl) edge[<->] (ra);
\end{tikzpicture}
\caption{Diagram of the main concepts. The target of each arrow is at least as expressive as its source.}
\label{fig:concepts}\vspace{-9mm}
\end{wrapfigure}
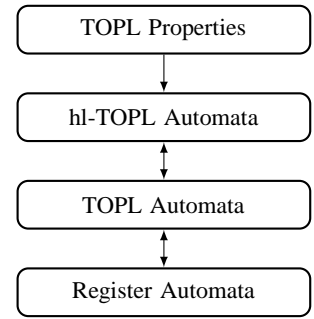
The aim of this work is to exploit the flexibility and the power of registers to address certain properties not {expressible} with other approaches
while, on the other hand,
making it easy for programmers to express properties of their code. 
More precisely, we start from register automata and extend them driven by typical properties required in real-world object-oriented systems.
This process results in the definition of two new classes of automata: \nt{
\begin{itemize}
\item {\emph{TOPL} automata}, which are low-level and are used for simplifying the formal correspondence with register automata;
\item {\emph{hl-TOPL} automata}, which are high-level and naturally express temporal properties about programs. 
\end{itemize}}\noindent
We moreover define the \boldemph{Temporal Object Property Language (TOPL)}, a formal language which maps directly onto hl-TOPL automata and is used for expressing runtime specifications. 
TOPL is a Java-programmer-friendly language where properties look like small Java programs that violate the desired program behaviour.
The hierarchy of presented concepts is depicted in Figure~\ref{fig:concepts}.

We complement and validate our theoretical results with a practical runtime verification tool for Java programs.
The tool can be used by programmers to rigorously express temporal properties about programs, which are then automatically checked by the system. 
Although the formal correspondence to register automata is completely hidden from the user, it provides a concrete automata-theoretic foundation which allows us to know formally the advantages and limitations of our technique, and also reuse results from the register automata literature.
%
%
Moreover, our tool can be tuned in terms of coverage, overhead and trace reporting by means of a number of parameters.

\paragraph{Contributions.}
\nt{This paper builds upon~\cite{our-fool2011}, which introduced the language of 
TOPL properties and drafted the corresponding automata.
Here we clarify the latter, provide a formal correspondence to automata 
over infinite alphabet, and devise and test a practical tool 
implementation.}
In summary, the contributions of the present work are:
\begin{asparaitem}
\item We introduce TOPL and hl-TOPL automata, two classes of abstract machines for verifying systems over infinite alphabets. We prove that both formalisms are equally expressive to register automata by constructing formal reductions between them.
The reductions allow us to transfer results from the register automata setting to ours (e.g.~decidability of language emptiness, language closures, etc.).
\item We define TOPL, a formal specification language designed for expressing program properties involving object interactions over time in a way that is familiar to object-oriented programmers.
We moreover present a formal semantics for TOPL, thus making it suitable for static and dynamic program analysis.
\item We implement a tool for automatically checking for violations of TOPL properties in Java programs at runtime.
A number of parameters are provided for tuning the precision of the system.
We furthermore report on experiments in which we ran our tool on large open-source projects. The results are encouraging: for example, we have found an interesting and previously unknown  concurrency bug in the DaCapo suite~\cite{dblp:conf/oopsla/dacapo}.
\end{asparaitem}

\section{Motivating Examples} 
\label{sec:motivating-examples}

Interaction among objects is at the core of the object-oriented paradigm.
Consider for example Java collections. A typical property one would want to state is
\eqquote{If one iterator modifies its collection then other iterators of the same collection become invalid, i.e.~they cannot be used further.}{q:concur-it}
The formalisation of the above constraint is non-trivial since it needs to keep track of {\em several objects} (at least two iterators and one collection) and their {\em interaction over time}.

A slightly more complex scenario is described in Figure~\ref{fig:first.java}.
Class {\tt CharArray} manipulates an array of chars, while
class {\tt Concat} concatenates two objects of type {\tt Str}.  Both classes implement the {\tt Str} interface.
Consider the case where {\tt Concat} is used for implementing a {\em rope}.\footnote{A rope is a data structure for efficiently storing and manipulating very long strings.}
The operations of a rope (e.g.~insert, concat, delete) may update its shape and the references to its root.
In this case we may have two or more collections {\em sharing} some elements. Hence, iterators operating on those different collections may invalidate each other. We need to modify~\eqref{q:concur-it}, increasing its complexity:
\eqquotee{If one iterator modifies its collection then other iterators of collections sharing some of its elements become invalid, i.e.~they cannot be used further.}{q:concur-it2}

\paragraph{On the need for re-binding.}
Let us now suppose we want to perform {\em taint checking} on input coming from a web form. What we want to check is the
property:
\eqquote{Any value introduced by the {\tt input()} method should not
  reach the {\tt sink()} method without first passing through the {\tt sanitizer()} method.}{q:concur-it3}
Although the property may seem simple, its difficulty can vary depending on the context.
Consider the case where
the input is constructed by concatenating strings from a web form, for example by using ropes implemented with class {\tt Concat}.
The number of user inputs, and therefore of concatenations, is not known a priori and is in general unbounded.
Consequently, we may end up having an unbounded number of tainted objects.
In a temporal specification, we would then need either one logical variable for each of them, or
the ability to {\em rebind} (or {\em update}) variables in the specification so that we can trace taint propagation.
For an unbounded number of objects, rebinding specification variables with
different values during the computation helps in keeping the specification finite.

\begin{figure}[t] 
\begin{minipage}{.4\linewidth}
\begin{verbatim}
interface Str {
  void set(int i, char c);
  char get(int i);
  int len();
  Itr iterator();
}
interface Itr {
  boolean hasNext();
  char next();
  void set(char c);
}
\end{verbatim}
\end{minipage}
\begin{minipage}{.45\linewidth}
\begin{verbatim}
class CharArray implements Str {
  char [] data;
  // ...
}
class Concat implements Str {
  Str s, t;
  public static Concat make(Str s, Str t) {
  /* ... */
  }
  // ...
}
\end{verbatim}
\end{minipage}
\caption{A first example: Java code.}
\label{fig:first.java}
\end{figure}

The need for rebinding of variables in the specification arises also in other contexts. For example,
when reasoning about the evolving shape of dynamically allocated data-structures.
Consider the following loop which uses a list:\\[1ex]
\centerline{
  $\texttt{\small while (l.next()!=null) \{ ... \}}$}\\[1ex]
If the list {\tt l}  contains a cycle, the loop will diverge.
Being a violation of a liveness property (termination), divergence cannot be observed at runtime in finite time and therefore it is harder to debug.
If we obtained the list by calling a third-party library, we
would want to check the property~\eqref{q:concur-it4} from the Introduction.
The {\em finite} encoding of such properties requires the ability to update the values of specification variables.

\section{TOPL Automata}\label{sec:topl_automata} 

We start by presenting some basic definitions.
We fix $V$ to be an infinite set of \emph{values}, with its members denoted by $v$, $u$ and variants. Given an arity $n$,
a \emph{letter} $\ell$ is an element $(v_1,\ldots,v_n)\in\Sigma$, where $\Sigma=V^n$ is the \emph{alphabet}. For $\ell=(v_1,...,v_n)$, we set the notation $\ell(i)=v_i$.
Given a size $m$,
we define the set of stores to be $S=V^m$. For a store $s=(u_1,...,u_m)$, we write $s(i)$ for $u_i$.
A \emph{register} $i$ is an integer from the set $\{1,...,m\}$ identifying a component of the store.

A \emph{guard} $g$ is a formula in a specified logic, interpreted over pairs of letters and stores;
we write $(s,\ell)\models g$ to denote that the store $s\in S$ and the letter $\ell\in\Sigma$ satisfy the guard~$g$, and we denote the set of guards by~$G$.
An \emph{action} $a$ is a small program which, given an input letter, performs a store update.
That is, the set of actions is some set $A\subseteq \Sigma\to S\to S$.

Given an alphabet $\Sigma=V^n$ and a (memory) size $m$, we shall define TOPL automata to operate on the set of labels $\Lambda=G\times A$, where $G$ and $A$ are given by:
\begin{align*}
G \;&::=\; {\sf eq}\,i\,j\ \mid\
{\sf neq}\,i\, j\ \mid\
 {\sf true}\ \mid\
G \mathrel{\sf and} G\\
A \;&::=\; {\sf nop}\ \mid\  {\sf set}\,i:=j\ \mid \ A;A
\end{align*}
with $i\in\{1,...,m\}$ and $j\in\{1,...,n\}$.
If $n=1$, then $({\sf eq}\,i)$ stands for $({\sf eq}\,i\,1)$; $({\sf neq}\,i)$ for $({\sf neq}\,i\,1)$; and $({\sf set}\,i)$ stands for $({\sf set}\,i:=1)$.
The guards are evaluated as follows.
\begin{itemize}
\item[] $(s,\ell) \models {\sf eq}\,i\,j$ \;\,\,
  if  $s(i)=\ell(j)$, \;\;
  $
  (s,\ell) \models {\sf true}$ \qquad\
  {always},
\item[]
$(s,\ell) \models {\sf neq}\,i\, j$ \;\
  if $s(i)\ne \ell(j)$, \;\;
  $(s,\ell) \models g_1 {\sf and}\, g_2$ \;\ if 
  $(s,\ell)\models g_1$ and $(s,\ell)\models g_2$.
\end{itemize}
The TOPL actions are built up from the empty action, ${\sf nop}(\ell)(s)=s$;
the assignment action, $({\sf set}\,i:=j)(\ell)(s)=s[i\mapsto \ell(j)]$
(where $s[i\mapsto v](k) = s(k)$ if $k\not=i$, and $v$ otherwise);
and action composition, $(a_1;a_2)(\ell)=a_1(\ell)\circ a_2(\ell)$.

We can now define our first class of automata.

\begin{definition}\label{def:sTOPL}
A \boldemph{TOPL automaton} with $m$ registers, operating on $n$-tuples, is a tuple ${\cal A}=\langle Q,q_0,s_0,\delta,F\rangle$ where:
\begin{compactitem}
\item $Q$ is a finite set of states, with initial one $q_0\in Q$ and final ones $F\subseteq Q$;
\item $s_0\in S$ is an initial store;
\item $\delta\subseteq Q\times\Lambda\times Q$ is a finite transition relation.
\end{compactitem}
\end{definition}

A \boldemph{configuration}~$x$ is a pair~$(q,s)$ of a state~$q$ and a store~$s$;
we denote the set of configurations by $X=Q\times S$.
The \emph{initial} configuration is $(q_0,s_0)$.
A configuration is \emph{final} when its state is final.
The configuration graph of a TOPL automaton ${\cal A}$ as above is a subset of $X\times\Sigma\times X$.
We write \ $x_1\stackrel{\ell}{\to}_{\cal A}x_2$ \ to mean that $(x_1,\ell,x_2)$ is in the configuration graph of~${\cal A}$
(we may omit the subscript if ${\cal A}$ is clear from the context).

\begin{definition}\label{def:cgraph}
Let ${\cal A}$ be a TOPL automaton.
The \boldemph{configuration graph} of ${\cal A}$ consists of exactly those configuration transitions \ $(q_1,s_1) \stackrel{\ell}{\to}_{\cal A} (q_2,s_2)$ \ for which there is a TOPL-automaton transition $(q_1,(g,a),q_2)\in\delta$ such that $(s_1,\ell)\models g$ and $a(\ell)(s_1)=s_2$.
\\
The \boldemph{language}~${\cal L}({\cal A})$ of ${\cal A}$ is the set of words that label walks from the initial configuration to some final one: \
${\cal L}({\cal A}) =
  \{\,\ell_1\ldots\ell_k\mid \text{$x_0$ initial,
  $x_k$ final,
  $\forall i\leq k.\,x_{i-1}\stackrel{\ell_i}\to_{\cal A} x_i$}\,\}$.
\end{definition}
A TOPL automaton is \boldemph{deterministic} when its configuration graph contains no two distinct transitions that have the same source~$x_1$ and are labeled by the same letter~$\ell$, that is,
$x_1\stackrel\ell\to x_2$ and $x_1\stackrel\ell\to x_3$ with $x_2\ne x_3$.

\begin{example}\label{ex:topl1}
\rg{Should we draw these two examples side by side and then comment only about one?}
\dd{yes. that's better. Or we can remove the first.}
Consider the language $\{\,abc\in V^3\mid\text{$a\ne c$ and $b\ne c$}\,\}$.
It is recognized by the following TOPL automaton with $2$~registers over the alphabet $\Sigma=V$. The values in the initial store~$s_0$ can be arbitrary.
\begin{itemize}
\item $Q=\{1,2,3,4\}$, $q_0=1$ and $F=\{4\}$;
\item $\delta=\{(1,({\sf true}, {\sf set}\,1),2)\}\cup\{
  (2,({\sf true}, {\sf set}\,2),3)\}\cup
  \{(3,({\sf neq}\,1\mathrel{\sf and}{\sf neq}\,2, {\sf nop}),4)\}$.
\end{itemize}
\end{example}

\begin{example}
For a more involved example, let us consider two languages over $\Sigma=V^2$:
\begin{align*}
{\cal L} &=
  \{\,(v_1,v_2)(v_2,v_3)(v_3,v_4)\ldots(v_{m-1},v_m)(v_m,v_1)
    \mid m>1 \;\land\; \forall i.\, v_i\in V\,\} \\
{\cal L}' &= \{\,w'\mid \exists w\in{\cal L}.\,\text{$w$ is a subsequence of $w'$}\,\}
\end{align*}
\end{example}

\noindent
\begin{wrapfigure}{r}{.25\linewidth}
\begin{minipage}{.25\linewidth}\vspace{-12mm}%
\begin{tikzpicture}\scriptsize
  \def\x{5}
  \def\y{1.2}
  \tikzset{vertex/.style={draw,circle,inner sep=1pt}}
  \tikzset{transition/.style={->,>=latex}}
  \tikzset{lbl/.style={right,inner sep=5pt}}
  \node[vertex] (start) at (0,0) [label=left:$q_0$] {};
  \node[vertex] (one) at (0,-1*\y) [label=left:$q_1$] {};
  \node[vertex,fill=darkred] (two) at (0,-2*\y) [label=left:$q_2$] {};
  \draw[transition] (-.25,0.25)--(start);
  \draw[transition] (start)--
    node[lbl]{${\sf set}\,1{:=}1;{\sf set}\,2{:=}2$} (one);
  \draw[transition] (start) .. controls +(-30:1cm) and +(30:1cm) ..
    node[right]{${\sf true},\,{\sf nop}$} (start);
  \draw[transition,thick] (one) .. controls +(-30:1cm) and +(30:1cm) ..
    node[lbl,pos=0.33]{${\sf eq}\,21,\,{\sf set}\,2{:=}2$}
    node[lbl,pos=0.67]{${\sf true},\,{\sf nop}$}
     (one);
  \draw[transition] (one)--
    node[lbl]{${\sf eq}\,21\,{\sf and\,eq}\,12$} (two);
  \draw[transition] (two) .. controls +(-30:1cm) and +(30:1cm) .. node[right]{$\;{\sf true},{\sf nop}$} (two);
\end{tikzpicture}
\end{minipage}\vspace{-11mm}
\end{wrapfigure}
(We say that $\ell_1\ldots\ell_m$ is a \emph{subsequence} of $\ell'_1\ldots\ell'_n$
  if there exists
    a strictly increasing $f:[m]\to[n]$ such that $f(i)=j$ only if $\ell_i=\ell'_j$.)
To account for chaining, as $\cal L$ requires,
  we will use two registers,
  one for~$v_1$ and one for the second component of the last seen letter.
To account for all subsequences, as ${\cal L}'$ requires,
  we will use nondeterminism.
In particular, the state~$q_1$ has \emph{two} loops,
  one guarded by $\sf true$ and another guarded by~${\sf eq}\,21$.

\paragraph{Relation to Register Automata.}
%
There is a natural connection between TOPL automata and \emph{Register Automata}~\cite{dblp:journals/tcs/kaminskif94,dblp:conf/mfcs/nevensv01}.
In particular,
register automata are TOPL automata with $n=1$ and labels from~$\Lambda_{\rm R}\subseteq\Lambda$,
where
\[
\Lambda_{\rm R} =
       \{\,({\sf fresh},{\sf set}\,i)\mid i\in\{1,...,m\}\,\}
  \cup \{\,({\sf eq}\,i,{\sf nop})\mid i\in\{1,...,m\}\,\}
\]
and ${\sf fresh}\equiv({\sf neq}\,1 \mathrel{\sf and} {\sf neq}\,2 \mathrel{\sf and} \cdots \mathrel{\sf and}  {\sf neq}\,m )$.\footnote{%
Here a register automaton corresponds directly to what in~\cite{dblp:conf/mfcs/nevensv01} is called a 1N-RA.}
In fact, we can show that the restrictions above are not substantial, in the sense that TOPL automata are reducible to register automata, and therefore equally expressive. 
In the following statement we use the standard injection $f:(V^n)^*\to V^*$ such that
$f({\cal L}({\cal A})) = \{\,v_1^1\ldots v_1^n\cdots v_k^1\ldots v_k^n \mid (v_1^1,\ldots,v_1^n)\cdots (v_k^1,\ldots,v_k^n)\in{\cal L}({\cal A})\,\}$. 

\begin{proposition}[TOPL to RA]\label{prop:topl-to-ra}
There exists an algorithm that, given a TOPL automaton~${\cal A}$, builds a register automaton~${\cal A}'$ such that ${\cal L}({\cal A}')=f({\cal L}({\cal A}))$.
If $\cal A$~has $m$~registers, $|\delta|$~transitions, $|Q|$~states, and works over $n$-tuples, then ${\cal A}'$~has $2m+1$ registers, $|\delta'|=O(n(2m)^{2m}|\delta|)$ transitions and  $O((2m)^m|Q|+|\delta'|)$~states.
\end{proposition}

\paragraph{High-level Automata} 
%
%
TOPL automata seem to be lacking the convenience one would desire for verifying actual programs. In particular, when writing down a monitor for a specific violation, one may naturally not want to specify all other possible behaviours of the program (which may, though, be of relevance to other monitors).
\rg{I understand this part.
But, I think, only because we discussed.}
In fact, program behaviours not relevant to the violation under consideration can be \emph{skipped}, ignored altogether.
A possible solution for the latter could be to introduce loops with empty guards and actions, with the hope to consume the non-relevant part of the program behaviour. However, such a solution would not have the desired effect: the empty loops could also consume input relevant to the monitored violation.

The above considerations lead us to introduce a new kind of automaton where inputs can be skipped.
That is, at each configuration $x$ of a such an automaton, if an input does not match any of the guards of the available transitions from $x$ 
the automaton will skip that input and examine the next one. In order to accommodate cases where we want specific transitions to happen consecutively, without skipping in between, we allow our automata to operate on sequences of letters, rather than single ones.

\begin{definition}\label{def:rollback}
A \boldemph{high-level TOPL automaton (hl-TOPL)} is a tuple ~${\cal A}=\langle Q,q_0,s_0,\delta,F\rangle$ where:
\begin{compactitem}
\item $Q$ is a finite set~$Q$ of states, with initial one $q_0\in Q$ and final ones $F\subseteq Q$;
\item $s_0\in S$ is an initial store;
\item $\delta\subseteq Q\times\Lambda^*\times Q$ is a finite transition relation.
\end{compactitem}
\end{definition}

Although the definition of the syntax of high-level automata is very similar to that of ordinary TOPL automata, their semantics is quite different.
A \boldemph{high-level configuration (hl-configuration)} is a pair~$(x,w)$ of a configuration~$x$ and a word~$w$;
we denote the set of hl-configurations by $Y=X\times\Sigma^*$.
We think of $w$ as \emph{yet to be processed}.
A hl-configuration is \emph{initial} when its configuration is the initial configuration; that is, it has the shape $((q_0,s_0),w)$.
A hl-configuration is \emph{final} when its state is final and its word is the empty word; that is, it has the shape $((q,s),\epsilon)$, where $q\in F$ and $\epsilon$~is the empty word.
The hl-configuration graph is a subset of $Y\times\Sigma^*\times Y$.
We write \ $y_1 \stackrel w\hookrightarrow_{\cal A} y_2$ \  to mean that $(y_1,w,y_2)$ is in the hl-configuration graph of~${\cal A}$.

The following concept simplifies the definition of the hl-configuration graph.
For each store $s$ and sequence of pairs $(g_i,a_i)$ of guards and actions ($i=1,\dots,d$), we construct a TOPL automaton
\[
{\cal T}\bigl(s,(g_1,a_1),\ldots,(g_d,a_d)\bigr)
\]
with set of states $\{0,\dots,d\}$, out of which $0$~is initial and $d$~is final, initial store $s$, and transitions $(i{-}1,(g_i,a_i),i)$ for each $i=1,\dots,d$.
Recall that, in this case, $\ell_1\ldots\ell_d$ is accepted by the automaton when there exist configurations $x_0,x_1,\dots,x_d$ such that $x_0=(0,s)$ and $x_{i-1}\stackrel{\ell_i}\to x_i$, for each $i=1,\dots,d$.
If the store of~$x_d$ is~$s'$ we say that the automaton can accept $\ell_1\ldots\ell_d$ with store~$s'$.

\begin{definition}\label{def:rcg}
The \boldemph{configuration graph} of a hl-TOPL automaton ${\cal A}$ consists of two types of transitions:
\begin{compactitem}
\item \emph{Standard transitions}, of the form \
  $((q_1,s_1),ww')\stackrel w\hookrightarrow((q_2,s_2),w')$,\\
 when there exists $(q_1,\bar\lambda,q_2)\in\delta$ such that ${\cal T}(s_1;\bar\lambda)$ can accept $w$ with store~$s_2$.
\item \emph{Skip transitions}, of the form \
  $(x,\ell w)\stackrel\ell\hookrightarrow(x,w)$,\\
when no standard transition starts from $(x,\ell w)$.
\end{compactitem}\vspace{1mm}
The \boldemph{language}~${\cal L}({\cal A})$ of ${\cal A}$ is the set of words that label paths from an initial hl-configuration to \hbox{a final one:\
$
{\cal L}({\cal A})=
  \{\,w_1\ldots w_k\mid
    \text{$y_0$~initial, $y_k$~final, }\forall i\leq k.\,
    y_{i-1}\stackrel{w_i}\hookrightarrow y_i\}
$.}
\end{definition}

\begin{remark}
Note that a TOPL automaton~${\cal A}$ can be technically seen as a high-level one with singleton transition labels. However, its language is in general different from the one we would get if we interpreted ${\cal A}$ as a high-level machine.
For example,
let ${\cal A}$ be the TOPL automaton consisting of one transition labelled with the guard ${\sf eq}\,1$, from the initial state to the final state.
The alphabet is $\Sigma=V$ and the initial store has one register containing value~$v$.
The language of ${\cal A}$ consists of one word made of one letter, namely~$v$.
On the other hand, because of skip transitions, the language of ${\cal A}$ seen as a hl-TOPL automaton consists of all words that contain the letter~$v$.
\end{remark}

\begin{example}
Consider the following hl-TOPL automaton with $2$~registers over the alphabet $\Sigma=V=\{A,B\}$.
\begin{compactitem}
\item $Q=\{1,2,3\}$, $q_0=1$ and $s_0=(A,B)$ and $F=\{3\}$,
\item $\delta$ consists of $\bigl(1,\bigl[({\sf eq}\,2,{\sf nop}),({\sf eq}\,1,{\sf nop}),({\sf eq}\,2,{\sf nop})\bigr],2\bigr)$ and $\bigl(1,\bigl[({\sf eq}\,1, {\sf nop})\bigr],3\bigr)$.
\end{compactitem}\noindent
The language of this is automaton consists of those words in which the first $A$ is not surrounded by two $B$s.
\end{example}

We next present transformations between the two different classes of automata we introduced. First,
we can transform TOPL automata to high-level ones by practically disallowing skip transitions: we obfuscate the original automaton ${\cal A}$ with extra transitions to a non-accepting sink state, in such a way that no room for skip transitions is left.

\begin{proposition}[TOPL to hl-TOPL]\label{prop:topl-to-rtopl}
There exists an algorithm that, given a TOPL automaton~${\cal A}$ with $|Q|$~states, at most $d$ outgoing transitions from each state, and guards with at most $k$~conjuncts, it builds a hl-TOPL automaton~${\cal A}'$ with $|Q|+1$ states and at most $(d+k^d)|Q|$ transitions such that ${\cal L}({\cal A})={\cal L}({\cal A}')$.
\end{proposition}

The converse is more difficult.
A TOPL automaton simulates a given hl-TOPL one by delaying decisions.
Roughly, there are two modes of operation:
(1)~store the current letter in registers for later use, and
(2)~simulate the configuration transitions of the original automaton.
The key insight is that Step~2 is entirely a static computation.
To see why, a few details about Step~1 help.

The TOPL automaton has registers to store the last few letters.
The states encode how many letters are saved in registers.
The states also encode a repartition function that records which TOPL register simulates a particular hl-TOPL register or a particular component of a past letter.
The repartition function ensures that distinct TOP registers hold distinct values.
Thus, it is possible to perform equality checks between hl-TOPL registers and components of the saved letters using only the repartition function.
Similarly, it is possible to simulate copying a component of a saved letter into one of the hl-TOPL registers by updating the repartition function.
Because it is possible to evaluate guards and simulate actions statically, the run of the hl-TOPL automaton can be completely simulated statically for the letters that are saved in registers.

\begin{proposition}[hl-TOPL to TOPL]\label{prop:rtopl-to-topl}
There exists an algorithm that, given a hl-TOPL automaton~${\cal A}$, builds a TOPL automaton~${\cal A}'$ such that
$ {\cal L}({\cal A}) = {\cal L}({\cal A}')$.
If ${\cal A}$~is over the alphabet~$V^n$ with $m$~registers, $|Q|$~states, and $|\delta|$~transitions of length~$\le d$,
then ${\cal A}'$ is over the alphabet $V$ with $m'=m+(d-1)n$ registers, $O(d^2 (m+1)^m|Q|)$ states, and $O(d^2 (m+1)^{(m+n)}|\delta|)$ transitions.
\end{proposition}

\begin{remark}
Although Propositions~\ref{prop:topl-to-rtopl} and \ref{prop:rtopl-to-topl} imply that hl-TOPL and TOPL automata are equally expressive, the transformations between them are non-trivial and substantially increase the size of the machines (especially in the hl-TOPL-to-TOPL direction).
This discrepancy is explained by the different goals of the two models: TOPL automata are meant to be easy to analyse, while high-level automata are meant to be convenient for specifying properties of object-oriented programs.
The runtime monitors implement the high-level semantics directly.
\end{remark}

Since both TOPL and hl-TOPL automata can be reduced to register automata, using known results for the latter~\cite{dblp:conf/mfcs/nevensv01} we obtain the following.


\begin{theorem}\label{th:main}
TOPL and hl-TOPL automata share the following properties.
\begin{compactenum}
\item The emptiness and the membership problems are decidable.
\item The language inclusion, the language equivalence and the universality problems are undecidable in general.
\item The languages of these automata are closed under union, intersection, concatenation and Kleene star.
\item The languages of these automata are not closed under complementation.
\end{compactenum}
\end{theorem}

\nt{The first point of Theorem~\ref{th:main} guarantees that monitoring with TOPL automata is decidable. On the other hand, by the second point, it is not possible to automatically validate refactorings of TOPL automata. Closure under regular operations, apart from negation, allows us to write specifications as negation-free regular expressions. The final point accentuates the difference between property violation and validation.}

\section{TOPL Properties}\label{sec:topl} 
\newcommand{\den}[1]{[\![#1]\!]}
\newcommand{\denG}[1]{\den{#1}_G}
\newcommand{\denA}[1]{\den{#1}_A}
\newcommand\pred{grd}

In this section we describe the user-level \emph{Temporal Object Property Language (TOPL)},
which provides a programmer-friendly way to write down hl-TOPL automata relevant to runtime verification.
The full syntax of the language was presented in~\cite{our-fool2011}.
Below we give the main ingredients and define the translation from the language to our automata.

A TOPL property comprises a sequence of \boldemph{transition statements}, of the form  
\[
\texttt{source -> target: label}
\]
where \text{source} and \texttt{target} are identifiers representing the states of the described automaton. The sequence of statements thus
represents the transition relation of the automaton.
Each property must include distinguished vertices \start and \error, which correspond to the initial and (unique) final states respectively.

The set of labels has been crafted in such a way that it captures the observable events of program executions. Observable events for TOPL properties are method calls and returns, called \boldemph{event ids},
along with their parameter values.
The set of event ids is given by
the grammar:
\rg{This isn't quite right.
$E$ is a set of patterns that will match (because of subtyping) sets of methods; the ids refer to concrete methods in these sets.
This is one reason why {\tt toplc} cannot compile a TOPL property independently on the program being observed.
It also shows we do something about inheritance.
It's not clear to me that it's good to hide this distinction, between patterns (sets of methods) and methods.}
\[
E \ ::= \ call\ m \mid ret\ m
\]
\nt{where $m$ belongs to an appropriate set of \boldemph{method names}}. Each method name has an \emph{arity}, which we shall in general leave implicit. 
The set $V_L$ of possible parameter values is a set of values specified by 
the programming language (e.g.~Java)
plus a dummy value $\bot$. The set of all values is $V=V_L\cup E$.

Labels of TOPL properties refer to registers via \boldemph{patterns}. 
A register  \Verb@v@ is called a
\boldemph{property variable} and has three associated patterns:
\begin{compactitem}
\item the uppercase pattern \Verb@V@ matches any value and writes it in the property variable \Verb@v@;
\item the lowercase pattern \Verb@v@ reads the value of the property variable \Verb@v@ and only matches that value; and
\item the negated lowercase pattern \Verb@!v@ reads the value of the
property variable \Verb@v@ and only matches different values.
\end{compactitem}
In addition, every element of $V$ acts as a pattern that matches only the value it denotes,
and a wildcard (\texttt{*}) pattern matches any value. 
\nt{The set of all patterns is denoted by  $Pat$.}
A transition label can take one of the three forms:
\[
  l \  ::= \ 
  call\;m(x_1, \ldots, x_{k})\  \mid\ 
  ret\;x := m\ \mid\
  x := m(x_1, \ldots, x_{k})
\]
where $x,x_1, \ldots, x_{k}\in Pat$. 
Note that the latter two forms are distinct\,--\,the last one incorporates a call and a matching return.
Finally, a TOPL property is \emph{well-formed} when it satisfies the conditions:
\begin{compactenum}[(i)]
\item each label must contain an uppercase value pattern at most once;
\item any use of a lowercase pattern (i.e. a read) must be preceded by a use of the corresponding uppercase pattern (i.e., a write) on all paths from \start.
\end{compactenum}
From now on we assume TOPL properties to be well-formed.

\paragraph{From TOPL to automata.}
We now describe how a TOPL property $P$ yields a corresponding hl-TOPL automaton $\A_P$.
First, if $n$ is the maximum arity of all methods in $P$, the alphabet of $\A_P$ will be:
\[
\Sigma_P = E \times V_L^{n+1}
\]
where the extra register is used for storing return values. Note that $\Sigma_P$ follows our previous convention of alphabets: it is a sub-alphabet of $\Sigma=V^{n+2}$.
For example, if $P$ has a maximal arity 5, the event $call\ m(a, b, c)$
would be understood as $(call\ m, \bot,  a, b, c, \bot, \bot)$ by
$\A_P$. Here the first component is the event id, the second is a
filler for the return value, the next three are the parameter values and the rest are
paddings which are used in order for all tuples to have the same length. The event $ret\ r = m$
would be understood as $(ret\ m, r, \bot, \bot, \bot, \bot, \bot)$ by
$\A_P$.

We include in $\A_P$ one register for each property variable in $P$ and,
in order to match elements from $V$, we include an extra register for each element mentioned by $P$ (this
includes all the method names of $P$). Each extra register contains a
specified value in the initial state of $\A_P$ and is never overwritten.
The rest of the registers in the initial state are empty. 
\nt{We write $Pat_P$ for the set of patterns of property variables appearing in $P$.}

We next consider how labels are translated.
The first two forms of label ($call$ and $ret$) describe observable events and are translated
into one-letter transitions in $\A_P$, while the latter form is translated into two-letter transitions.
Let $\{1,\ldots,N\}$ be the set of registers of $\A_P$. We define three functions:
$reg : Pat_P\to (N\cup\{\bot\})$ associates a register to each pattern (with $reg(\texttt{*})=\bot$), while $\pred: Pat_P\times N\to G$ and $act: Pat_P\times N\to A$ give respectively the guard and action correspoding to each pattern and register. We set:
\[
\pred(x,j) = \begin{cases}
{\sf true} & \text{if }  x=\mathtt{V} \\
{\sf eq}\,reg(\mathtt{v})\,j  &\text{if }   x=\mathtt{v} \\
{\sf neq}\,reg(\mathtt{v})\,j  &\text{if } x=  \mathtt{!v}\\
{\sf eq}\,reg(x)\,j  &\text{if }x\in V \\
{\sf true} &\text{if }x=\texttt{*}
\end{cases}
\quad
act(x,j) = \begin{cases}
{\sf set}\,{\it reg}({\tt v}):=j  & \text{if }  x=\mathtt{V} \\
{\sf nop} &\text{if }   x=\mathtt{v} \\
{\sf nop} &\text{if } x=  \mathtt{!v}\\
{\sf nop} &\text{if }x\in V \\
{\sf nop} &\text{if }x=\texttt{*}
\end{cases}
\]
We can now interpret labels of $P$ into labels of $\A_P$.
For each a label $l$ of $P$, we define its translation 
$
\den{l} = [(\denG{l}, \denA{l})]
$,
where $\denG-$ and $\denA-$ are given as follows.
\begin{align*}
\denG{l} &=\begin{cases}
\pred(m,1)\,\mathsf{and}\,\pred(x_1,3)\, \mathsf{and}\ldots\mathsf{and}\, \pred(x_k,k{+}2) & \text{if } l=  call\;m(x_1, \ldots, x_k) 
\\
pred(m,1) \;\mathsf{and}\; {\sf eq}\,reg(x)\, 2 & \text{if }  l = ret\;x := m
\end{cases} 
\\
\denA{l} &= \begin{cases}
act(x_1,3)\; \mathsf{and}\ldots\mathsf{and}\; act(x_k,k{+}2) & \text{if } l=  call\;m(x_1, \ldots, x_k) 
\\
act(x,2) & \text{if }  l = ret\;x := m
\end{cases}
\end{align*}
Finally, for the label $x := m(x_1, \ldots, x_k)$,
observe its right-hand-side refers to a call, while its left-hand-side refers to a return. We take this label to
mean that $m$ is called with parameters matching $x_1, \ldots, x_k$
and returns a value matching $x$, \emph{and no event is observed in the meantime}. 
This is because an intermediate call, for instance a
recursive call, could disconnect the method call and the return value.
Thus, this label translates into a transition of length two:
\[
\den{x := m(x_1, \ldots, x_k)}
=
\den{call\;m(x_1, \ldots, x_k)}\, \den{ret\;x := m}
\]

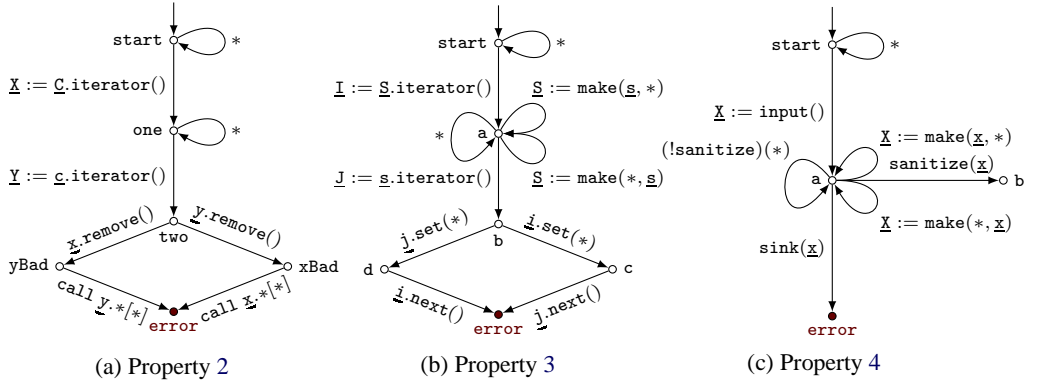
\begin{figure}[t]
\hbox{%
\begin{minipage}{.345\linewidth}\centering
\begin{tikzpicture}\scriptsize
  \def\x{1.5}
  \def\y{1.2}
  \tikzset{vertex/.style={draw,circle,inner sep=1pt}}
  \tikzset{transition/.style={->,>=latex}}
  \node[vertex] (start) at (0,0) [label=left:\texttt{start}] {};
  \node[vertex] (one) at (0,-1*\y) [label=left:\texttt{one}] {};
  \node[vertex] (two) at (0,-2*\y) [label=below:\texttt{two}] {};
  \node[vertex] (xBad) at (1*\x,-2.5*\y) [label=right:\texttt{xBad}] {};
  \node[vertex] (yBad) at (-1*\x,-2.5*\y) [label=left:\texttt{yBad}] {};
  \node[vertex,fill=darkred] (error) at (0,-3*\y) [label=below:\textcolor{darkred}{\texttt{error}}] {};
  \draw[transition] (0,0.5)--(start);
  \draw[transition] (start)--node[left]{$\pattern X:=\pattern C.\mathtt{iterator}()$} (one);
  \draw[transition] (start) .. controls +(30:1cm) and +(-30:1cm) .. node[right]{$*$} (start);
  \draw[transition] (one) .. controls +(30:1cm) and +(-30:1cm) .. node[right]{$*$} (one);
  \draw[transition] (one)--node[left]{$\pattern Y:=\pattern c.\mathtt{iterator}()$} (two);
  \draw[transition] (two) -- node[sloped,above]{$\pattern y.\mathtt{remove}()$} (xBad);
  \draw[transition] (two)--node[sloped,above]{$\pattern x.\mathtt{remove}()$} (yBad);
  \draw[transition] (xBad)--node[sloped,below]{$\;\;\;\mathtt{call}\;\pattern x.{*}[*]$} (error);
  \draw[transition] (yBad)--node[sloped,below]{$\mathtt{call}\;\pattern y.{*}[*]\;\;\;$} (error);
\end{tikzpicture}
(a) Property~\ref{q:concur-it}
\end{minipage}
\begin{minipage}{.345\linewidth}\centering
\begin{tikzpicture}\scriptsize
  \def\x{1.5}
  \def\y{1.2}
  \tikzset{vertex/.style={draw,circle,inner sep=1pt}}
  \tikzset{transition/.style={->,>=latex}}
  \node[vertex] (start) at (0,0) [label=left:\texttt{start}] {};
  \node[vertex] (one) at (0,-1*\y) [label=left:\texttt{a}] {};
  \node[vertex] (two) at (0,-2*\y) [label=below:\texttt{b}] {};
  \node[vertex] (xBad) at (1*\x,-2.5*\y) [label=right:\texttt{c}] {};
  \node[vertex] (yBad) at (-1*\x,-2.5*\y) [label=left:\texttt{d}] {};
  \node[vertex,fill=darkred] (error) at (0,-3*\y) [label=below:\textcolor{darkred}{\texttt{error}}] {};
  \draw[transition] (0,0.5)--(start);
  \draw[transition] (start)--node[left]{$\pattern{I} := \pattern{S}.\mathtt{iterator()}$} (one);
  \draw[transition] (start) .. controls +(30:1cm) and +(-30:1cm) .. node[right]{$*$} (start);
  \draw[transition] (one) .. controls +(130:1.2cm) and +(-130:1.3cm) .. node[left]{$*$} (one);
  \draw[transition] (one) .. controls +(60:1cm) and +(0:1cm) .. node[anchor=south west,pos=.25]{$\pattern{S} := \mathtt{make(\pattern{s}, *)}$} (one);
  \draw[transition] (one) .. controls +(-60:1cm) and +(0:1cm) .. node[anchor=north west,pos=.25]{$\pattern{S} := \mathtt{make(*,\pattern{s})}$} (one);  
  \draw[transition] (one)--node[left]{$\pattern{J} := \pattern{s}.\mathtt{iterator()}$} (two);
  \draw[transition] (two) -- node[sloped,above]{$\pattern{i}.\mathtt{set(*)}$} (xBad);
  \draw[transition] (two)--node[sloped,above]{$\pattern{j}.\mathtt{set(*)}$} (yBad);
  \draw[transition] (xBad)--node[sloped,below]{$\;\;\pattern{j}.\mathtt{next()}$} (error);
  \draw[transition] (yBad)--node[sloped,below]{$\pattern{i}.\mathtt{next()}\;\;$} (error);
\end{tikzpicture}
(b) Property~\ref{q:concur-it2}
\end{minipage}
\begin{minipage}{.345\linewidth}\centering
\begin{tikzpicture}\scriptsize
  \def\x{1.5}
  \def\y{1.2}
  \tikzset{vertex/.style={draw,circle,inner sep=1pt}}
  \tikzset{transition/.style={->,>=latex}}
  \node[vertex] (start) at (0,0) [label=left:\texttt{start}] {};
  \node[vertex] (one) at (0,-1.5*\y) [label=left:\texttt{a}] {};
  \node[vertex] (blind) at (1.5*\x,-1.5*\y) [label=right:\texttt{b}] {};
  \node[vertex,fill=darkred] (error) at (0,-3*\y) [label=below:\textcolor{darkred}{\texttt{error}}] {};
  \draw[transition] (0,0.5)--(start);
  \draw[transition] (start)--node[left]{$\pattern{X} := \mathtt{input()}$} (one);
  \draw[transition] (start) .. controls +(30:1cm) and +(-30:1cm) .. node[right]{$\mathtt{*}$} (start);
  \draw[transition] (one) .. controls +(130:1.2cm) and +(-130:1.3cm) .. node[anchor=south east,pos=.35]{$\mathtt{(!sanitize)(*)}\!\!$} (one);
  \draw[transition] (one) .. controls +(0:1cm) and +(60:1cm) .. node[anchor=south west,pos=.55]{$\pattern{X} := \mathtt{make(\pattern{x}, *)}$} (one);
  \draw[transition] (one) .. controls +(0:1cm) and +(-60:1cm) .. node[anchor=north west,pos=.55]{$\pattern{X} := \mathtt{make(*,\pattern{x})}$} (one);  
  \draw[transition] (one)--node[above,pos=.65]{$\mathtt{sanitize(\pattern{x})}$} (blind);
  \draw[transition] (one)--node[left]{$\mathtt{sink(\pattern{x})}$} (error);  
\end{tikzpicture}
(c) Property \ref{q:concur-it3}
\end{minipage}%
}\caption{TOPL formalisations of the example properties from Section~\ref{sec:motivating-examples}.}\label{fig:examples}
\end{figure}

\paragraph{Examples.} \label{sec:examples} 

\autoref{fig:examples} displays the formal versions of the first three properties that are discussed in \autoref{sec:motivating-examples}.

(a)~This example illustrates how multiple related objects are tracked.
In state {\tt two}, the property tracks all pairs of two iterators $x$~and~$y$ for the same collection~$c$.
If $x.{\tt remove}()$ is called, then state {\tt yBad} becomes active, which precludes further use of $y$'s methods.
State {\tt xBad} is symmetric.

(b)~This example illustrates how chaining of values is tracked, while at the same time tracking multiple related objects.
Recall that Property~\eqref{q:concur-it2} refers to the code in \autoref{fig:first.java} (on page~\pageref{fig:first.java}).
In state~{\tt a}, the iterator~$i$ refers to the string~$s$ or some substring of~$s$.
In state~{\tt b}, the itrerator~$j$ refers to~$s$.
As opposed to the previous property, the two iterators $i$~and~$j$ are not necessarily for the same collection, but rather for a collection and one of its sub-collections.
This property does not refer to the Java standard library, which does not implement ropes.
There exist, however, several independent libraries that follow the pattern in \autoref{fig:first.java} (e.g.\  {\small\url{http://ahmadsoft.org/ropes/}}).

(c)~This example illustrates sanitization of values, in addition to chaining.
In state~{\tt a}, the property keeps track of the tainted object~$x$.
An object is \emph{tainted} if it comes from a specific input method or was made from tainted objects, and was not sanitized.
A tainted object must not be sent to a {\tt sink}.

Of course, the input of the TOPL compiler is not in graphical form.
Below we include the actual representation for a property of type~(c) without the sanitization option.
It specifies actual methods that provide tainted inputs, make tainted objects out of tainted objects, and constitute sinks.\footnote{\nt{%
Note that, to ease the task of writing TOPL properties, we have included a \texttt{prefix}
directive: $\mathtt{prefix}\; p$ produces from every
method name $m$, an extra name $pm$; it further
produces, from any transition involving $m$, a similar transition
involving $pm$.}}
We refer the reader to~\cite{our-fool2011} for more example properties.

\smallskip
\begin{Verbatim}[fontsize=\footnotesize,commandchars=\\\{\}]
property Taint
  prefix <javax.servlet.http.HttpServletRequest>
  prefix <java.lang.String>
  prefix <java.sql.Statement>
  start -> start:       *
  start -> tracking:    \pattern{X} := *.getParameter[*]
  tracking -> tracking: *
  tracking -> tracking: \pattern{X} := \pattern{x}.concat(*)
  tracking -> tracking: \pattern{X} := *.concat(\pattern{x})
  tracking -> error:    *.executeQuery(\pattern{x})
\end{Verbatim}

\section{Implementation and Experiments} \label{sec:implementation}

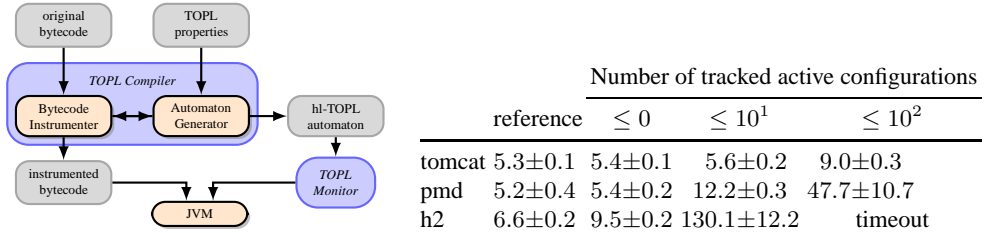
\begin{figure}[t]
\begin{center}
\begin{tabular}{ll}
\begin{tikzpicture}[node distance=12pt and 2pt, auto, >=latex]
\tiny
\tikzstyle{system}=[rectangle,
                                draw=blue!60,
                                fill=blue!20,
                                inner sep=0.1cm,
                                rounded corners=8pt,
                                style=thick]
\tikzstyle{program}=[rectangle,
                                  draw=black,
                                  text width=30pt,
                                  text centered,
                                  fill=orange!20,
                                  inner sep=0.1cm,
                                  rounded corners=5pt,
                                  style=thick,
                                  drop shadow]
\tikzstyle{data}=[rectangle,
                            draw=gray!70,
                            text width=30pt,
                            text centered,
                            fill=gray!30,
                            inner sep=0.1cm,
                            rounded corners=5pt,
                            style=thick]
\node[program]
  (instrumenter) {Bytecode Instrumenter};
\node[program, right=15pt of instrumenter]
  (genautomaton) {Automaton Generator};
\node[above=1pt] at ($(instrumenter.north)!.5!(genautomaton.north)$) (topllabel) {\emph{TOPL Compiler}};
\begin{pgfonlayer}{background}
  \node[system, fit = (topllabel) (instrumenter) (genautomaton)] (TOPLC) {};
\end{pgfonlayer}
\node[above=of TOPLC] (top) {};
\node[data] at (instrumenter |- top)
  (classes) {original bytecode};
\node[data] at (genautomaton |- top)
  (properties) {TOPL properties};
\node[right=of TOPLC,text width=50pt] (right) {};
\node[data] at (right |- genautomaton)
  (automaton) {hl-TOPL automaton};
\node[below=of TOPLC] (bot) {};
\node[data] at (instrumenter |- bot)
  (instrclasses) {instrumented bytecode};
\node[text width=20pt, text centered] at (automaton |- bot)
  (checkerlabel) {\emph{TOPL Monitor}};
\begin{pgfonlayer}{background}
  \node[system, fit = (checkerlabel)] (checker) {};
\end{pgfonlayer}
\node at ($(instrclasses)!.5!(checker)$) (classdummy) {};
\node[program, below=5pt of classdummy]
  (jvm) {JVM};

\path[thick, ->]
(classes) edge (instrumenter)
(genautomaton) edge (instrumenter)
(instrumenter) edge (genautomaton)
(properties) edge (genautomaton)
(genautomaton) edge (automaton)
(instrumenter) edge (instrclasses);

\draw[thick, ->]
(instrclasses) -| ($(jvm.north)-(5pt,0)$);
\draw[thick, ->]
(checker) -| ($(jvm.north)+(5pt,0)$);

\draw[thick, ->]
(automaton) -- (checker);
(\end{tikzpicture}
&
\begin{tabular}[b]{@{}lr@{}lr@{}lr@{}lr@{}l}
  &&
  & \multicolumn{6}{c}{Number of tracked active configurations} \\ \cmidrule{4-9}
& \multicolumn{2}{c}{reference}
  &\multicolumn{2}{c}{$\le0$}
  &\multicolumn{2}{c}{$\le10^1$}
  &\multicolumn{2}{c}{$\le10^2$}
\\ \midrule
tomcat
  & $5.3$ & $\pm0.1$
  & $5.4$ & $\pm0.1$
  & $5.6$ & $\pm0.2$
  & $9.0$ & $\pm0.3$
\\
pmd
  & $5.2$ & $\pm0.4$
  & $5.4$ & $\pm0.2$
  & $12.2$ & $\pm0.3$
  & $47.7$ & $\pm10.7$
  \\
h2
  & $6.6$ & $\pm0.2$
  & $9.5$ & $\pm0.2$
  & $130.1$ & $\pm12.2$
  & \multicolumn{2}{c}{timeout}
  \\
\end{tabular}
\end{tabular}
\caption{{\bf Left:} Architecture of the TOPL tool. {\bf Right:} Experimental Results.
  Times are in seconds, averaged over $10$~runs \nt{(not in convergence mode)}.}
\label{fig:architecture}
\end{center}
\end{figure}

The TOPL tool\footnote{\url{http://rgrig.github.com/topl}} checks at runtime whether Java programs violate TOPL properties.
It consists of a compiler and a monitor (see \autoref{fig:architecture}, left).

Given the bytecode of a Java project and several TOPL properties, the compiler produces instrumented bytecode and a hl-TOPL automaton.
An instrumented method emits a call event, runs the original bytecode, and then emits a return event.
Emitting an event is encoded by a call to the method {\tt check(Event)} of the monitor.
The {\tt Event} structure contains an integer identifier and an array of {\tt Object}s.
The identifier is unique for each site from which the method {\tt check} is called.
The compiler achieves two tasks that are interdependent: instrumenting the bytecode, and translating properties into an automaton.
The instrumentation could be done on all methods of the Java project, but this would lead to high runtime overhead.
Instead, the compiler instruments only the methods that are mentioned by the TOPL properties to be checked.
Conversely, the translation of properties into automata depends on the Java project's code.
To see why, consider a transition guarded in a property by the method name pattern~$m$.
The compiler instruments all the methods whose (fully qualified) names match the pattern~$m$, and all the methods that override methods whose names match the pattern~$m$, thus taking into account inheritance.
All these instrumented methods emit events with identifiers from a certain set of integers, which depends on the inheritance structure of the Java project.
The method name pattern~$m$ is essentially compiled into a set of integer event identifiers.

The monitor is an interpreter for the hl-TOPL automaton that the compiler produces.
Its implementation closely follows the semantics from \autoref{sec:topl_automata}.
For example, the monitor maintains a set of active configurations, which are those reachable by a path labeled by the events seen so far.
\rg{`Active configurations' deserves a formal definition, rather than the informal `For example$\ldots$'.}
There are, however, several differences.
First, the number of active configurations is not bounded in theory, but a bound may be enforced in practice.
Monitoring becomes slower as the number of active configurations increases.
As a pragmatic compromise, the user may impose an upper bound, thus trading soundness for efficiency.
That is, if the user imposes a bound then monitoring is faster, but property violations may be missed
(on the other hand, a reported violation of a property is always a real violation).
Second, the implementation includes several optimizations.
For example, the guards produced by method name patterns, which require the current event id to be from a certain set of integers, is implemented as a hashtable lookup rather than as a linear search, as the formal semantics would suggest.
Third, the implementation saves extra information in order to provide friendlier error messages.
For instance, the user may ask the TOPL monitor to save and report the path taken in the configuration graph, or full call stack traces for each event.

\paragraph{Experimental Results.}\label{sec:results} 
We measured the overhead on the test suite DaCapo~\cite{dblp:conf/oopsla/dacapo}, version~9.12.
DaCapo is a collection of automated tests that exercise large portions of code from open-source projects and the Java standard libraries.
DaCapo itself has been used for many experiments by the research community.
Hence, we did not expect to find any bugs, but aimed instead at measuring the overhead.
We checked two types of properties with TOPL\null.
First, properties that express correct usage of the standard Java libraries.
Second, properties that express temporal constraints which we extracted from the code comments of three open-source projects (H2, PMD, Tomcat) included in DaCapo.
 H2 is a database server for which we checked properties on the calling order of some interface methods. For example, a client should not attempt to ask for a row from a cursor unless the latter has been previously advanced. PMD looks for bugs, dead code and other problems in Java
code. One of the five properties we checked is
``Only if a scope replies that it knows a name, it can be asked for that name's definition''.
Tomcat is a highly concurrent servlet server.
Servlets are Java programs running in a webserver, extracting data from {\tt ServletRequest}s and sending data to {\tt ServletResponse}s. A response has two associated incoming channels: a stream and a writer. They should not be both used concurrently. But the servlet, before forwarding the response, must call {\tt flush} on the stream, on the writer, or on the response itself. This is one of the properties we checked.
Interestingly, while experimenting with Tomcat, TOPL discovered
a concurrency bug (a data race) in the  DaCapo's infrastructure which would manifest sometimes as  {\tt null} dereference.

Although our tool is not currently optimised,
we measured both time and space overhead. It turns out that space
overhead is
negligible, below the variance caused by the randomness of garbage collection.
Thus, we only report on time overhead, in \autoref{fig:architecture} and \autoref{table:experiments}.
The relative overhead is meaningful only if the reference runtime is not close to~$0$, and
this is most distinctively the case for test eclipse whose runtime is over $10\,\rm s$.
The (geometric) average overhead in that case is $\times1.5$ with $\le3$ active configurations, and $\times1.6$ with $\le10$ active configurations.
Figure~\ref{fig:architecture} shows the effect of tuning the active configurations in terms of overhead.
All experiments were performed on an Intel i5 with $4$~cores at $3.33\rm\,GHz$
with  $4\rm\,GiB$ of memory,
running Linux 2.6.32 and Java VM~1.6.0\_20.

\begin{table}[t]
\begin{center}
\begin{tabular}{|l|r||r|r||r|r||r|r||r|r|} \hline
  & original  & \multicolumn{2}{|c||}{HasNext} & \multicolumn{2}{c||}{UnsafeIterator} &\multicolumn{2}{|c||}{UnsafeMapIterator} & \multicolumn{2}{|c|}{UnsafeFileWriter}
  \\ \hline
  & & st=3 & st=10 & st=3 & st=10 & st=3 & st=10 & st=3 & st=10
  \\ \hline \hline
 avrora  & 8.1 & 27.8 & 60.5 & 163.3 & 323.1 & 194.5  & 179.9 & 8.3 & 5.9
 \\ \hline
 batik     &  1.2  & 18.1 &    3.0 & 3.8 &  3.8  & 3.1 &  3.3 & 1.3  &  1.2
 \\ \hline
eclipse   &  17.4  &  24.2 &  24.0 & 30.9 &  41.7 & 27.2 &  28.0 & 22.9 &  22.8
\\ \hline
fop    &   0.3   &  0.9 &  1.9 & 3.5  &  3.6   & 2.7 &   2.7 & 0.3 &   0.3
\\ \hline
h2     &    6.2  &  5.9 &  6.8   & 8.3    & 20.0  & 13.5 & 11.2   & 6.4 & 6.0
\\ \hline
jython    &  1.9  & 19.8 &  46.1 &  81.5   & 83.0  &  62.8 &       62.7    & 1.9 & 1.8
\\ \hline
luindex     &  0.8   & 0.8 &   0.8    & 0.8 & 0.9  & 1.0  &   0.9 & 0.8 & 0.9
\\ \hline
lusearch    &  1.5  &  1.5 &   1.5 & 15.0 &  16.0    & 13.8  &  12.8    & 1.5 & 1.7
\\ \hline
pmd        &   3.1    & 19.9 & 42.6  &  93.5 & 240.3   & 102.6 &       105.6  & 3.2 &           3.3
\\ \hline
sunflow    &  3.9   & 3.8 &  3.9  &  4.0 & 3.8    & 3.9 &        3.9     & 3.9 &         4.3
\\ \hline
tomcat     &   2.5    & 4.2 & 8.3   & 22.9 & 50.9   & 30.0 &        31.0   & 2.6 &          2.7
\\ \hline
xalan      &    1.5   &  14.5      & 7.1 & 425.0  & 360.9 &  272.0   & 276.5 &  1.5 & 1.2
\\ \hline
\end{tabular}
\end{center}
\caption{Experiment on small properties (taken from [25]) run on the DaCapo benchmarks
  (in convergence mode).   HasNext checks that no iterator is advanced without first enquiring hasNext. UnsafeIterator checks that no iterator is advanced after the iterated collection has been modified. UnsafeMapIterator checks that no iterator on keys/values of a map is advanced after the map has been updated. UnsafeFileWriter checks that no file is written to after it was closed.
Column original gives the running times (in seconds) for  projects without instrumentation of Java standard libraries. The other columns report instrumented runs, with a maximum of 3 and 10 active configurations.}
\label{table:experiments}
\end{table}

\section{Related work.}

{\it JavaMOP\/}~\cite{dblp:journals/sttt/meredithjgcr12} and {\it Tracematches\/}~\cite{dblp:conf/oopsla/allanachklmsst05} are based on slicing:
A slice is a projection of a word over a finite alphabet; different slices are fed, independently, to machines that handle finite alphabets.
Tracematches use regular expressions to specify recognisers over finite alphabets.
JavaMOP supports several other logics, via a plugin mechanism, and slices are assigned categories, which can be match\slash fail or taken from some other set.
Because slices are analyzed independently, it not possible to express examples such as \eqref{q:concur-it4}~and~\eqref{q:concur-it3}, which use an unbounded number of register assignments.

{\it Quantified Event Automata (QEA)}~\cite{dblp:conf/fm/barringerfhrr12} extend the slicing mechanism of JavaMOP with the goal of improving expressivity.
Similarly to TOPL automata, QEAs have guards and assignments, which can be arbitrary predicates and transformations of the memory content respectively.
In contrast, our automata impose specific restrictions, which follow the expressive power of RAs.
In addition, QEAs introduce quantifiers, which can be seen as a way to impose a hierarchy on slices.
Systems based on machines with parametric transition rules, such as {\it RuleR\/}~\cite{dblp:journals/logcom/barringerrh10}, {\it LogScope\/}~\cite{barringer2010formal} and {\it TraceContract\/}~\cite{dblp:conf/fm/barringerh11}, are related to QEAs and are also very similar in spirit to the TOPL approach.
RuleR is tuned towards high expressivity and in particular can handle context-free grammars with parameters, which go beyond the reach of TOPL.
By comparison, TOPL automata seem a simpler formalism, and this paper demonstrates how they are closely related to standard automata-theoretical models.

{\it QVM\/}~\cite{arnold:2008} is a runtime monitoring approach tailored to deployed systems.
It achieves high efficiency by 
being carefully implemented inside a Java virtual machine,
checking properties involving a single object, and being 
able to tune its overhead on-the-fly.
On the other hand, TOPL is designed for aiding the programmer during development and testing,
and therefore focusses instead on providing a precise and expressive language for specifying temporal properties.
For instance, TOPL can express properties involving many objects.
Both QVM and TOPL let the programmer tune the overhead\slash coverage balance.
{\it ConSpec\/}~\cite{DBLP:journals/entcs/AktugN08} is a language used to describe security policies.
Although ConSpec automata have a countable number of states, they are deterministic and therefore cannot express the full range of TOPL properties.

From the techniques used mostly for static verification of object-oriented programs, {\it typestates\/}~\cite{strom1986} are probably the most similar to TOPL.
A modular static verification method for typestate protocols is introduced in~\cite{dblp:conf/oopsla/bierhoffa07}.
The specification method is based on linear logic, and relations among objects in the protocol are checked by a tailored system of permissions.
Similarly,~\cite{deline2004,dblp:conf/sigsoft/BierhoffA05} provide a means to specify typestate properties that belong to a single object.
The specified properties are reminiscent of contracts or method pre/post-conditions and can deal with inheritance.
In~\cite{dblp:conf/issta/FinkYDRG06} the authors present sound verification techniques for typestate properties of Java programs, which we envisage that can be fruitfully combined with the TOPL paradigm.
Their approach is divided into several stages each employing its own verifier, with progressively higher costs and precisions. 
Every stage focuses on verifying only the parts of the code that previous stages failed to verify.

A specification language for interface checking aimed at C programs, called {\it SLIC\/}, is introduced in~\cite{ball2002}.
SLIC uses non-determinism to encode universal quantification of dynamically allocated data and allows for complex code in the automaton transitions; while
TOPL specifications naturally express universally quantified properties over data structures and, for effectiveness reasons, there is a limit on the actions performed during automaton transitions.
Simple SLIC specifications are verified by  the SLAM verifier~\cite{dblp:conf/cav/ballr01}.

Similar investigations have been pursued by the functional programming community.
In~\cite{disney2011} contracts are used for expressing legal traces of programs in a functional language with references.
The contracts specify traces as regular expressions over calls and returns, thus resembling our automata, albeit in quite a different setting.
The specifications are function-centered and, again, capturing inter-object relations seems somewhat tricky.

Finally, as demonstrated in previous sections, TOPL automata are a variant of register automata~\cite{dblp:journals/tcs/kaminskif94,dblp:conf/mfcs/nevensv01}, themselves a thread in an extensive body of work on automata over infinite alphabets (see e.g.~\cite{dblp:conf/csl/segoufin06}).
RAs form one of the most well-studied paradigms in the field, with numerous extensions and variations (e.g.~\cite{dblp:conf/lics/bojanczykmssd06,dblp:journals/tcs/bjorklunds10,dblp:conf/csl/segoufin06}).

\paragraph{Acknowledgements.} We acknowledge funding from EPSRC H011749 (Distefano and Grigore) and RAEng research fellowships (Distefano and Tzevelekos).
We thank the anonymous reviewers, for their suggestions significantly improved the paper.

\bibliographystyle{plain}
\bibliography{safety}

\newpage\appendix
\section{Proofs}\label{sec:proofs}

TOPL automata are are a superclass of register automata.
It is relatively harder to establish that TOPL automata are at most as expressive as register automata.
There are three main ingredients in the proof.
First, tuples $(v_1,\ldots,v_n)$ are unpacked into words $v_1\ldots v_n$; correspondingly, transitions $q\to q'$ become paths $q\to q_1\to\cdots\to q_{n-1}\to q'$.
Second, register~$i$ is simulated by register~$r(i)$, where $r\in[m]\to[m]$ is encoded in the state, such that values are not repeated in registers, as required by register automata.
Third, locally fresh values are written to an extra register, because register automata never ignore such values.

On its own, the second ingredient is the same as the proof of Kaminski and Francez~\cite{dblp:journals/tcs/kaminskif94} that their M-automata are equally expressive to register automata.
In the proof below, however, the first two ingredients (unpacking tuples and ensuring that values do not repeat) are blended together to improve the bounds.
The encoding of the function~$r$ leads to a $m^m$ expansion, in the worst case.
It turns out that a similar function is needed for unpacking tuples, which would lead to another similar expansion if the first two ingredients would not be mixed.

Consider the label $({\sf eq}\,1\,2,{\sf set}\,1:=1)$, with the alphabet~$V^2$.
When tuples are unpacked, it would be tempting to replace it by the two labels, $({\sf true},{\sf set}\,1)$ and $({\sf eq}\,1,{\sf nop})$, one for each component of the tuple.
But, this would be incorrect, as the second component should be compared to the old value of register~$1$.
The solution is to add extra registers and a function similar to~$r$ that keeps track of which register simulates which register.

\begin{oneshot}{Proposition~\ref{prop:topl-to-ra} (TOPL to RA)}
There exists an algorithm that, given a low-level TOPL automaton~${\cal A}$, builds a register automaton~${\cal A}'$ such that ${\cal L}({\cal A}')=f({\cal L}({\cal A}))$, where $f\bigl((v_1,\ldots,v_n)\bigr)=v_1\ldots v_n$ extends to languages as follows
\[f({\cal L}({\cal A})) = \{\,f(v_1)\ldots f(v_k)\mid v_1\ldots v_k\in{\cal L}({\cal A})\,\}\]
If $\cal A$~has $m$~registers, $|\delta|$~transitions, $|Q|$~states, and works over $n$-tuples, then ${\cal A}'$~has $2m+1$ registers, $|\delta'|=O(n(2m)^{2m}|\delta|)$ transitions, and  $O((2m)^m|Q|+|\delta'|)$~states.
\end{oneshot}
\begin{proof}
Each state~$q$ of~${\cal A}$ is encoded by $(2m)^m$ states $(q,r)$ of~${\cal A}'$, one for each $r\in[m]\to[2m]$.
While in a state~$(q, r)$, register~$r(i)$ of~${\cal A}'$ simulates register~$i$ of~${\cal A}$.
Each transition $(q,(g,a),q')$ in~${\cal A}$ is encoded by paths
\begin{align*}
(q_0,r_0)\xrightarrow{(g_1,a_1)}(q_1,r_1)\xrightarrow{(g_2,a_2)}\cdots\xrightarrow{(g_n,a_n)}(q_n,r_n)
\end{align*}
with $q_0=q$ and $q_n=q'$.
For each~$r_0$ there are at most $(2m+1)^n$ such paths, because each $(q_j,r_j)$ has at most $2m+1$ successors, as described below.

The following invariants are maintained along the path.
The values held by registers~$r_0([m])$ are not changed -- they simulate the values that ${\cal A}$~holds in state~$q$.
The guards~$g_j$ use only the current letter and the registers~$r_0([m])$.
The values held by registers~$r_n([m])$ in state~$(q_n,r_n)$ are the values that ${\cal A}$~holds in state~$q'$.
In intermediate states~$(q_j,r_j)$, the registers~$r_j([m])$ hold a mixture of the values held by~$r_0([m])$ and those held by~$r_n([m])$ in state~$(q_n,r_n)$.
A more precise description of the mixture follows:
Let the action~$a_{(\le j)}$ be obtained from the original action~$a$ by filtering out assignments $({\sf set}\,i':=j')$ with $j<j'$.
Then registers~$r_j([m])$ in state~$(q_j,r_j)$ simulate what the registers of~${\cal A}$ would be in state~$q'$ if $a$ would be replaced by~$a_{(\le j)}$.

\smallskip
The invariants mentioned in the previous paragraph are maintained by constructing the guards~$g_j$ and the actions~$a_j$ as follows.

Let $I_g$ be the set $\{\,i\mid\text{${\sf eq}\,i\,j$ occurs in~$g$}\,\}$ of registers that must equal the current component~$j$; similarly, let $I'_g=\{\,i\mid\text{${\sf neq}\,i\,j$ occurs in~$g$}\,\}$.
If $|r_0(I_g)|\ge2$ or $|r_0(I_g)\cap r_0(I'_g)|\ge1$, then $(q_{j-1},r_{j-1})$~has no successor.
If $r_0(I_g)=\{i\}$, then $(q_{j-1},r_{j-1})$ has exactly one successor, and the guard~$g_j$ is ${\sf eq}\,i$.
If $r_0(I_g)=\emptyset$, then $(q_{j-1},r_{j-1})$ has $|[2m]-r_0(I'_g)|+1$ successors with the guards {\sf fresh} and, respectively, ${\sf eq}\,i$ for $i\in[2m]-r_0(I'_g)$.

The action~$a_j$ and the function~$r_j$ are computed from the original action~$a$, the previous function~$r_{j-1}$, and from the guard~$g_j$, which is described in the previous paragraph.
Let $I_a=\{\,i\mid\text{$a$~writes component~$j$ to register~$i$}\,\}$.
First a target register in~${\cal A}'$ is picked, and then the saving of component~$j$ in registers~$I_a$ is simulated.
The target~$k\in[2m]$, which is needed only if $I_a\ne\emptyset$, is picked as follows:
\begin{align*}
k=\begin{cases}
i &\text{if $g_j$ is ${\sf eq}\,i$} \\
\min \bigl([2m]-r_0([m])-r_{j-1}([m]-I_a)\bigr) &\text{if $g_j$ is {\sf fresh}}
\end{cases}
\end{align*}
The action~$a_j$ depends on~$g_j$ and on~$I_a$.
\begin{align*}
a_j=\begin{cases}
  {\sf nop} &\text{if $g_j$ is ${\sf eq}\,i$, or $I_a=\emptyset$} \\
  {\sf set}\,k &\text{if $g_j$ is {\sf fresh}, and $I_a\ne\emptyset$}
\end{cases}
\end{align*}
Finally, the repartition function is updated to reflect that registers~$I_a$ are now simulated by~$k$.
\begin{align*}
r_j(i) =\begin{cases}
  k &\text{if $i\in I_a$} \\
  r_{j-1}(i) &\text{if $i\notin I_a$}
\end{cases}
\end{align*}

At this point the labels have the form $({\sf eq}\,i,{\sf nop})$ or $({\sf fresh},{\sf set}\,i)$ or $({\sf fresh},{\sf nop})$.
Only the latter is disallowed by the definition of register automata.
It can be handled by adding one register, without significantly increasing the number of transitions.
First, each label $({\sf fresh},{\sf nop})$ is transformed into $({\sf fresh},{\sf set}\,2m+1)$.
Second, for each transition labeled $({\sf fresh},a)$, we add a parallel transition labeled $({\sf eq}\,2m+1,a)$.
\qed
\end{proof}

\begin{oneshot}{Proposition~\ref{prop:topl-to-rtopl} (TOPL to hl-TOPL)}
There exists an algorithm that, given a low-level TOPL automaton~${\cal A}$ with $|Q|$~states, at most $d$~transitions outgoing of each state, and guards with at most $k$~conjuncts, builds a high-level TOPL automaton~${\cal A}'$ with $|Q|+1$ states and at most $(d+k^d)|Q|$ transitions such that ${\cal L}_\rho({\cal A}')={\cal L}({\cal A})$.
\end{oneshot}
\begin{proof}
For the automaton~${\cal A}$ compare the low-level configuration graph with the high-level configuration graph.
Each low-level transition $x_1\to^\ell x_2$ corresponds to several high-level standard transitions $ (x_1,\ell w)\hookrightarrow^\ell (x_2,w)$, for all $w$.
The high-level graph, however, also has skip transitions $(x,\ell w)\hookrightarrow^\ell (x,w)$ for configurations~$x$ that have no outgoing standard transitions.
Thus, the low-level language and the high-level language would agree if all low-level configurations would have at least one outgoing transition.

We obtain ${\cal A}'$ from~${\cal A}$ by adding unit transitions that do not change the low-level language, but ensure that all low-level configurations have an outgoing transition.
First we add a special stuck state~$q_{\rm stuck}$.
Then, for each original state~$q$, we list the guards $g_1$,~$g_2$, \dots,~$g_d$ of the outgoing transitions.
The configuration $(q,s)$ has no outgoing transition for some~$\ell$ when $(s,\ell)\not\models g_i$ for $i\in[k]$.
So, we synthesize a guard~$g$ that holds exactly in this situation.
Informally, we want to add a transition from~$q$ to~$q_{\rm stuck}$ with the guard $g=\lnot g_1\land\ldots\land\lnot g_d$ and the action ${\sf nop}$.
Such a guard is not expressible in the TOPL guard logic.
However, we can negate the simple guards ${\sf eq}$ and ${\sf neq}$, we can use distributivity to put $g$ in disjunctive normal form, and we can simulate disjunction by parallel transitions.
Thus, if each $g_i$ contains up to $k$ simple conjuncts, we add at most $k^d$ transitions from state~$q$.
\qed
\end{proof}

\begin{oneshot}{Proposition~\ref{prop:rtopl-to-topl} (hl-TOPL to TOPL)}
There exists an algorithm that, given a high-level TOPL automaton~${\cal A}_\rho$, builds a low-level TOPL automaton~${\cal A}$ such that
$ {\cal L}_\rho({\cal A}_\rho) = {\cal L}({\cal A})$.

If ${\cal A}_\rho$ is over the alphabet~$V_\rho^n$ with $m_\rho$~registers, $|Q_\rho|$~states, and $|\delta_\rho|$~transitions of length~$\le d$,
then ${\cal A}$ is over the alphabet $V_\rho$ with $m=m_\rho+(d-1)n$ registers, $O(d^2 (m+1)^m|Q_\rho|)$ states, and $O(d^2 (m+1)^{(m+n)}|\delta_\rho|)$ transitions.
\end{oneshot}
\begin{proof}
The set of states of~${\cal A}$ is \[Q=Q_\rho\times\{0,\ldots,d-1\}\times\{0,\ldots,d-1\}\times\bigl([m]\pmap[m]\bigr)\]
When the high-level configuration
\[ \bigl(q_\rho,(v_1,\ldots,v_{m_\rho})\bigr),\ell_0\ldots\ell_{h-1}\]
of~${\cal A}_\rho$ is reached, one of the low-level configurations
\[ (q_\rho,h,k,r),\;(u_1,\ldots,u_m) \]
of~${\cal A}$ is reached.
Here $h$~is the length of the remembered history, while $k$,~$r$, $u_1$, \dots,~$u_m$ vary subject to the following constraining invariants:
\begin{itemize}
\item the stores of~${\cal A}$ are injective: $u_i \ne u_j$ if $i\ne j$
\item for $i\in[m_\rho]$, the value of register~$i$ of~${\cal A}_\rho$ is stored in register $r(i)$ of~${\cal A}$; that is, $ u_{r(i)} = v_i $
\item for $0\le k'<h$ and $\ell_{k'}=(v'_1,\ldots,v'_n)$ and $1\le i'\le n$, the $i'$th component~$v'_{i'}$ of letter~$\ell_{k'}$ is stored in register $r(\iota(k+k',i'))$ of~${\cal A}$, where $\iota(k'',i')=m_\rho+(k''\bmod d)n+i'$; that is,
  \[u_{r(\iota(k+k',i'))}=v'_{i'}\]
\item $r$ is undefined for register slots that are reserved for storing letters but are currently unused; namely, $r(\iota(k+k',i'))$ is undefined for all $i'$~and~$k'$ such that $1\le i'\le n$ and $h\le k'<d$
\end{itemize}

\smallskip
We consider in turn each state $q=(q_\rho,h,k,r)$.
Its outgoing transitions are determined by the outgoing transitions of~$q_\rho$.

We must cater for two situations: Either more letters are arriving and
we should simulate storing them in the queue and possibly make a state
transition, or no more letters are arriving and we should simulate
emptying the queue to see if we end up in an accepting state.

Crucially, enough information is available to statically simulate receiving all the letters in the queue.
To see this, consider a transition in ${\cal A}_\rho$ out of ~$q_\rho$,
\[ (q_\rho, [(g_0,a_0),\ldots,(g_{d'-1},a_{d'-1})], q'_\rho)
  \qquad\text{with $d'\leq h$} \]
We can assume $d'\leq h$ as only transitions short enough to be
evaluated with the received letters may be taken.

Consider an assignment $({\sf set}\,i:=j)$ that appears in~$a_{k'}$.
Suppose that the distribution of values just before this assignment is given by~$r'$.
This means that the register~$i$ of~${\cal A}_\rho$ is simulated by $r'(i)$, and that the $j$th component of letter~$\ell_{k'}$ is simulated by $r'(\iota(k+k',j))$.
After the assignment is executed, the distribution of values is given by
\[r''(i') = \begin{cases}
  r'(\iota(k+k',j)) & \text{if $i'=i$} \\
  r'(i') & \text{otherwise}
\end{cases}\]
Thus, it is possible to statically find the register distribution function after
each of the $d'$~steps.

Let us write $r_{k'}$ for the distribution function just before step~$k'$; in particular, $r_0=r$, where $r$~is given by the state~$q$ of~${\cal A}$.
Suppose now that $g_{k'}$ contains the conjunct $({\sf eq}\,i\,j)$.
We can evaluate this conjunct statically by checking whether $r_{k'}(i)=r_{k'}(\iota(k+k',j))$.
Similarly, we can evaluate $({\sf neq}\,i\,j)$ by checking whether $r_{k'}(i)\ne r_{k'}(\iota(k+k',j))$ as the store is injective.
Thus we can evaluate all guards $g_0$,~$g_1$, \dots,~$g_{d'-1}$.

If one of the guards is not true, we know that the transition would
not be taken, if the queue were to be emptied. If all guards are true,
we know that the resulting state is $(q'\rho, h-d',( k+d')\bmod
d,r_{d'})$.
If $h-d' > 0$ then we carry on simulating. If $h-d' = 0$ then we note
if $q'_\rho$ is final in ${\cal A}_\rho$.

If we find that no transitions are taken, then we must compute the
result of a skip transition. This can also be done statically by
incrementing~$k$, decrementing~$h$ and
 replacing the distribution function by one that is undefined at
 $\iota(k,i')$ for $i'\in[n]$. If $k-1 > 0$, we carry on simulating and
 if $k-1 = 0$, we note
if $q_\rho$ is final in ${\cal A}_\rho$.

 If any of these simulations thus notices a final state,
then $(q_\rho, h, k, r)$ is final in ${\cal A}$.

Now to handle the case of more incoming letters, we need to add transtions.
We treat three cases: Firstly, if the queue is not full ($h < d-1$) then we
simply store the current letter in the queue. Secondly, if the queue
is full ($h=d-1$) and
none of the transitions in ${\cal A}_\rho$ out of $q_\rho$ has maximal
length (length $d$), then we can statically determine if we need to simulate a
high-level standard transition or a skip transition and what the resulting states
would be. Thirdly, if the queue is full, and there is a transition of
maximal length, then we need to dynamically look at the current letter
to determine if that transition is taken and where it leads. Note
that this also determines whether a skip transition should be simulated.

{\it Case $h<d-1$}.
In this case all outgoing transitions of~$q$ simply record the current letter $\ell=(v_1,\ldots,v_n)$.
To maintain the injectivity of stores, only those components of~$\ell$ that are not already in some register must be stored.
We consider $(m+1)^n$ distinct situations: each of the $n$~components may be in one of the $m$~registers, or it may be fresh.
Such a situation is described by a function $p\in[n]\pmap[m]$.
We add to~${\cal A}$ a transition
\[ (q_\rho,h,k,r),\;(g_p,a_p),\;(q_\rho,h+1,k,r_p)\]
The guard~$g_p$ is constructed such that it ensures we are indeed in a situation described by~$p$; the action~$a_p$ stores the fresh values of~$\ell$ somewhere outside of $r([m])$; the function~$r_p$ records where the fresh values were stored and where the existing values already were.

The guard $g_p$ is constructed as follows.
If $p(j)$ is undefined, which means that $v_j$ should be fresh, then $g_p$ contains conjuncts $({\sf neq}\,i\,j)$ for $i\in[m]$.
If $p(j)$ is defined, which means that $u_{p(j)}=v_j$, then $g_p$ contains the conjunct $({\sf eq}\,p(j)\,j)$.
These are all the conjuncts of~$g_p$.

We now fix some injection $\sigma$ from the set $\{v_j\mid\text{$p(j)$ undefined}\}$ of fresh values to the set $[m]-r([m])$ of unused registers.
The action $a_p$ contains an assignment $({\sf set}\,\sigma(v_j):=j)$ for each $j$ where $p(j)$~is undefined.
Also
\[ r_p(\iota(k',i')) = \begin{cases}
  p(i')  & \text{if $k'=k+h$ and $p(i')$ defined} \\
  \sigma(v_{i'}) & \text{if $k'=k+h$ and $p(i')$ undefined} \\
  r(\iota(k',i')) &\text{otherwise}
\end{cases}\]

{\it Case $h=d-1$, no outgoing transitions of length $d$}.
At this point the values in the $m$~registers of~${\cal A}$ are enough
to decide whether to simulate a standard or a skip transition
of~${\cal A}_\rho$.
The construction above is used to add transitions which save the
current letter. However, each such transition is added a number of
times, one for each outgoing transition in ${\cal A}_\rho$. The
targets of these transitions are modified to reflect the effect of
taking the transition. This can be determined statically as described
above. Specifically, it is known at this point if any of the transitions
can be taken. We only add the ones that would (we cannot determine a
target for the others anyway). If no transitions can be taken, we
simulate a standard transition. This is again done by storing the
current letter, but we also drop the letter at the front of the queue
(by incrementing~$k$ and decrementing~$h$ and
 replacing the distribution function by one that is undefined at $\iota(k,i')$ for $i'\in[n]$).

{\it Case $h=d-1$, ${\cal A}_\rho$ has an outgoing transition of length $d$}.
At this point the values in the $m$~registers of~${\cal A}$ together
with the current letter are needed to decide whether to simulate a
standard or a skip transition of~${\cal A}_\rho$.

As in the case above, we can statically evaluate all transitions of
length shorter than $d$ and add transitions for them. But we cannot
add the encoding of a skip transition because it should only be taken
if no standard transitions are. For each standard transition of length
$d$
\[ (q_\rho, [(g_0,a_0),\ldots,(g_{d-1},a_{d-1})], q'_\rho), \]
we can statically evaluate up to the point right before the final
guard $g_{d-1}$. Thus we can add an automata transition to ${\cal A}$
with $(g'_{d-1}, a'_{d-1})$ on it and target $(q'_\rho, 0, 0, r')$, where
$g'_{d-1}$ and $a'_{d-1}$ are versions of $g_{d-1}$ and $a_{d-1}$
modified to refer to $r_{d-1}$ and $r'$ is a version of $r_d$ which is
undefined on all indices pointing into the queue.

The guard $g'_{d'-1}$ is constructed as follows:
For all conjuncts ${\sf eq}\,i\,j$ in $g_{d'-1}$, $g'_{d'-1}$ contains ${\sf eq}\,r_{d'-1}(i)\,j$ and
for all conjuncts ${\sf neq}\,i\,j$ in $g_{d'-1}$, $g'_{d'-1}$ contains ${\sf neq}\,r_{d'-1}(i)\,j$.

Finally, we must also simulate a skip transition, but only to be taken in case none of the other transitions are.
That is, if any of the short transitions are taken, we have no skip
transition. If none of the short transitions are taken, we construct a
guard that is true if none of the final guards for the maximal length
transitions are. Guards to ensure this are generated using the same
construction employed in the proof of \autoref{prop:topl-to-rtopl}. These are then combined with the
construction in the previous case.
\qed
\end{proof}


\end{document}